\documentclass[preprint,showkeys,showpacs,tightenlines,fleqn,prd]{revtex4}   
\usepackage{graphicx}  
\begin{document}        
\preprint{PNU-NTG-02/2007}  
\preprint{PNU-NuRI-02/2007}  
\title{Kaon semileptonic decay ($K_{l3}$) form factors\\ from the instanton  
  vacuum}      
\author{Seung-il Nam}  
\email{sinam@pusan.ac.kr}  
\affiliation{Department of  
Physics and Nuclear Physics \& Radiation Technology Institute (NuRI),  
Pusan National University, Busan 609-735, Republic of Korea}   
\author{Hyun-Chul Kim}  
\email{hchkim@pusan.ac.kr}  
\affiliation{Department of  
Physics and Nuclear Physics \& Radiation Technology Institute (NuRI),  
Pusan National University, Busan 609-735, Republic of Korea}   
\date{\today}  
\begin{abstract}  
We investigate the kaon semileptonic decay ($K_{l3}$) form  
factors within the framework of the nonlocal chiral quark model  
($\chi$QM) from the instanton vacuum, taking into account the effects  
of flavor SU(3) symmetry breaking.  We also consider the problem of  
gauge invariance arising from the momentum-dependent quark mass in   
the present work.  All theoretical calculations are carried out  
without any adjustable parameter, the average instanton size  
($\rho\sim1/3$ fm) and the inter-instanton   
distance ($R\sim1$ fm) having been fixed.  We also show that the  
present results satisfy the Callan-Treiman low-energy theorem as well  
as the Ademollo-Gatto theorem.  Using the $K_{l3}$ form   
factors, we evaluate relevant physical quantities.  It turns out that  
the effects of flavor SU(3) symmetry breaking are essential in  
reproducing the kaon semileptonic form factors.  The present results  
are in a good agreement with experiments, and are compatible with  
other model calculations.    
\end{abstract}   
\maketitle  
\section{introduction}  
It is of great importance to understand semileptonic decays of kaons  
($K_{l3}$), since it plays a significant role in determining the CKM  
matrix element $|V_{us}|$  
precisely~\cite{Cabibbo:1963yz,Leutwyler:1984je}.  Furthermore, it   
provides a simple phenomenological basis for testing the breaking of   
flavor SU(3) symmetry: In exact flavor SU(3) symmetry, the kaon  
semileptonic form factor $f_+(0)$ becomes unity.  The  
Ademollo-Gatto~\cite{Ademollo:1964sr} theorem asserts that of second  
order are the corrections of flavor SU(3) symmetry breaking to the  
form factors of vector currents at zero momentum transfer ($q^2=0$).   
However, when Goldstone bosons are involved, the Ademollo-Gatto  
theorem must be modified: While the singlet part of flavor SU(3)  
symmetry breaking preserves flavor SU(3) symmetry, it breaks chiral    
SU(3)$\times$SU(3) symmetry.  Langacker and Pagels showed that the  
corrections of flavor SU(3) symmetry breaking appear to first order  
due to the presence of that singlet  
part~\cite{Langacker:1973nf,BaceCornwell,Gasser:1984ux}.    
The effect of flavor SU(3) symmetry breaking on the kaon semileptonic  
decay form factor is known to be around $3\sim5\%$, which is rather  
small.    
  
The well-known soft-pion Callan-Treiman~\cite{Callan:1966hu} theorem  
connects the ratio of the pion and kaon decay constants to the  
semileptonic form factors of the kaon at $q^2=m_K^2-m_\pi^2$  
(Callan-Treiman point).  Any chiral quark model should satisfy the  
Callan-Treiman theorem with flavor SU(3) symmetry breaking.  
Experimentally, there are a certain amount of data to judge  
theoretical calculations~\cite{Yao:2006px,Chounet:1971yy}.  Thus, the  
kaon semileptonic decay form factor provides a basis to examine the   
validity and reliability of any theoretical theory and model for  
hadrons.  
  
There has been a great number of theoretical work:   
chiral perturbation theory  
($\chi$PT)~\cite{Gasser:1984ux,Bijnens:1994me}, lattice QCD   
(LQCD)~\cite{Becirevic:2004ya,Tsutsui:2005cj}, a Dyson-Schwinger  
method~\cite{Ji:2001pj,Kalinovsky:1996ii}, constituent quark  
models~\cite{Isgur:1975hu,Choi:1998jd,Choi:1999nu}, and so on.     
In the present work, we will investigate the $K_{l3}$ form factor  
within the framework of the nonlocal chiral quark model ($\chi$QM)  
derived from the instanton vacuum.  We will consider the leading order  
in the large $N_c$ expansion and flavor SU(3) symmetry breaking  
explicitly.  The meson-loop corrections, which are of $1/N_c$  
order, are neglected.  The model has several virtues: All relevant QCD  
symmetries are satisfied within the model, and there are only two 
parameters: The average size of instantons ($\rho\sim1/3$ fm) and 
average inter-instanton distance ($R\sim1$ fm), which can be 
determined by the internal constraint such as the saddle-point   
equation~\cite{Shuryak:1981ff,Diakonov:1983hh,Diakonov:1985eg}.  These   
values for $\rho$ and $R$ have been supported in various LQCD  
simulations recently~\cite{Chu:vi,Negele:1998ev,DeGrand:2001tm}.     
There is no further adjustable parameter in the model.  
  
As being discussed previously, since the effects of flavor SU(3)  
symmetry breaking are essential in the present work, we employ the   
modified low-energy effective partition function with flavor SU(3)  
symmetry  
breaking~\cite{Musakhanov:1998wp,Musakhanov:2001pc,Musakhanov:2002vu}.    
This partition function extends the former one derived in the chiral  
limit~\cite{Diakonov:1983hh,Diakonov:1985eg}.  It has been proven that  
the partition function with flavor SU(3) symmetry breaking is very  
successful in describing the low-energy hadronic properties such as  
various QCD condensates, magnetic susceptibilities, meson distribution  
amplitudes, and so  
on~\cite{Kim:2004hd,Nam:2006sx,Nam:2006ng,Ryu:2006bf,Musakhanov:2002xa}.   
However, the presence of the nonlocal interaction between quarks and  
pseudo-Goldstone bosons breaks the Ward-Takahashi identity for   
N\"other currents.  Since the kaon semileptonic decay form factors  
involve the vector current, we need to deal with this problem.  While  
Ref.~\cite{Pobylitsa:1989uq} proposed a systematic way as to how the  
conservation of the N\"other current is restored, one has to handle  
the integral equation.  Refs.~\cite{Musakhanov:2002xa,Kim:2004hd}  
derived the light-quark partition function in the presence   
of the external gauge fields.  With this gauged partition function, it  
was shown that the low-energy theorem for the transition from  
two-photon state to the vacuum via the axial anomaly was  
satisfied~\cite{Musakhanov:2002xa}.  Moreover, the magnetic   
susceptibility of the QCD vacuum and the meson distribution amplitudes     
were obtained sucessfully~\cite{Kim:2004hd,Nam:2006sx}.  Thus, in the  
present work, we will investigate the kaon semileptonic decay  
($K_{l3}$) form factors, using the gauged low-energy effective   
partition function from the instanton vacuum with flavor SU(3)  
symmetry breaking explicitly taken into account.    
  
We sketch the present work as follows: In Section II, we briefly  
explain the general formalism relevant for studying the $K_{l3}$ form  
factor.  In Section III, we introduce the nonlocal chiral quark model  
from the instanton vacuum.  In Section IV, the numerical results are   
discussed, and are compared with those of other works.  The final  
Section is devoted to summarize the present work and to draw   
conclusions.    
\section{Semileptonic kaon decay}  
\begin{figure}[h]  
\begin{center}  
\includegraphics[width=10cm]{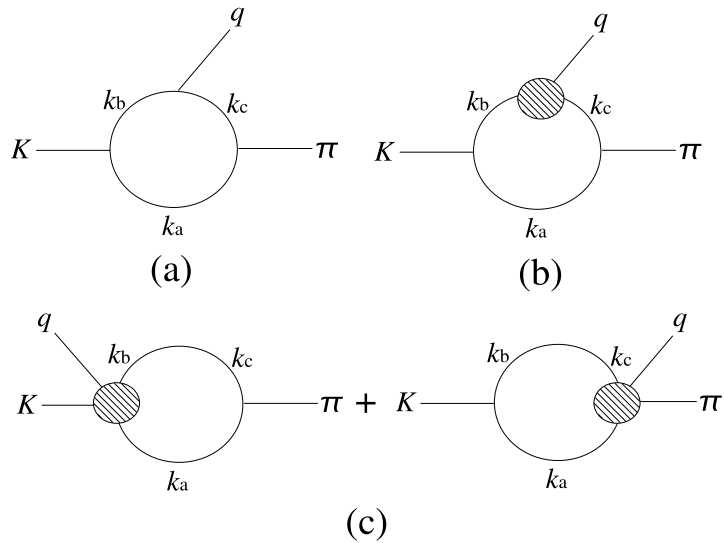}  
\end{center}  
\caption{Schematic diagrams for the kaon semileptonic decay form factor. We  
consider the contributions from the local (a) and  
nonlocal vector-quark vertices (b) and nonlocal vector-quark-meson  
vertices (c). Here, we define the relevant momenta as follows:  
$k_a=k-p/2-q/2$, $k_b=k+p/2-q/2$ and $k_c=k+p/2+q/2$, where $k$, $p$  
and $q$ stand for the loop-integral variable, initial kaon and  
vector-field momenta, respectively.}   
\label{fig0}  
\end{figure}

In the present work, we are interested in the following kaon semileptonic  
decays ($K_{l3}$) in two different isospin channels:   
\begin{eqnarray}  
  \label{eq:DEF1}  
K^+(p_K)&\to&\pi^0(p_{\pi})\,l^+(p_l)\,\nu_l(p_{\nu}):\,K^{+}_{l3},  
\nonumber\\  
K^0(p_K)&\to&\pi^-(p_{\pi})\,l^+(p_l)\,\nu_l(p_{\nu}):\,K^{0}_{l3},  
\end{eqnarray}  
where $l$ and $\nu_l$ stand for the leptons (either the electron or  
the muon) and neutrinos.  The relevant diagrams for the $K_{l3}$ form  
factor are depicted in Figure~\ref{fig0} in which we define the  
momenta for the particles involved.    
The nonlocal contributions in Figure~\ref{fig0}(b) and \ref{fig0}(c)  
arise from the gauged effective chiral action that will be discussed 
in Section III.    
  
The decay amplitude ($T_{K\to l\nu\pi}$) can be expressed as  
follows~\cite{Bijnens:1994me}:    
\begin{equation}  
  \label{eq:T}  
T_{K\to l\nu\pi}=\frac{G_F}{\sqrt{2}}\sin\theta_c\left[w^{\mu}(p_l,p_{\nu})  
F{\mu}(p_K,p_{\pi})\right],   
\end{equation}  
where $G_F$ is the well-known Fermi constant: 
$G_F=1.166\times10^{-5}\,{\rm{GeV}}^{-2}$.  $\theta_c$ denotes the 
Cabibbo angle. We define, respectively, the weak leptonic  
and hadronic matrix elements ($w^{\mu}$ and $F_{\mu}$) with the 
$\Delta S=1$ vector current ($j^{su}_{\mu}$) as: 
\begin{eqnarray}  
  \label{eq:DEF2}  
w^{\mu}(p_l,p_{\nu})&=&\bar{u}(p_{\nu})\gamma^{\mu}(1-\gamma_5)v(p_l),  
\\  
\label{eq:DEF3}  
F_{\mu}(p_K,p_{\pi})  
&=&c\langle \pi(p_\pi)|j^{su}_{\mu}|K(p_K)\rangle  
=c\langle \pi(p_\pi)|\bar{\psi}\gamma_{\mu}\lambda^{4+i5}  
\psi|K(p_K)\rangle   
\nonumber\\  
&=&(p_K+p_{\pi})_{\mu}f_{l+}(t)+(p_K-p_{\pi})_{\mu}f_{l-}(t),  
\end{eqnarray}   
where $c$ is the isospin factor, and set to be unity and $1/\sqrt{2}$  
for $K^0_{l3}$ and $K^+_{l3}$, respectively.  The matrix  
$\lambda^{4+i5}$ denotes the combination of the two Gell-Mann   
matrices, $\left(\lambda^4+i\lambda^5\right)/2$, for the  
relevant flavor in the present problem.  The $\psi$ denotes the quark    
field.  The momentum transfer is defined as  
$Q^2=(p_K-p_{\pi})^2\equiv{-t}$.     
  
$f_{l\pm}$ represent the vector form factors with the corresponding  
lepton $l$ ($P$-wave projection).  Alternatively, the form factor  
$F_{\mu}(p_K,p_{\pi})$ can be expressed in terms of the scalar form 
factor ($f_{l0}$, $S$-wave projection) and the vector form factor 
$f_{l+}$ defined as follows:    
\begin{equation}  
  \label{eq:scalar}  
F_{\mu}(p_K,p_{\pi})=f_{l+}(t)(p_K+p_{\pi})_{\mu}  
+\frac{(m^2_{\pi}-m^2_K)(p_K-p_{\pi})_{\mu}}{t}  
\left[f_{l+}(t)-f_{l0}(t)\right].    
\end{equation}  
Hence, the $f_{l0}$ can be written as the linear combination of  
$f_{l+}$ and $f_{l-}$:    
\begin{equation}  
  \label{eq:swave}   
  f_{l0}(t)=f_{l+}(t)+\left[\frac{t}{m^2_K-m^2_{\pi}}\right]f_{l-}(t).  
\end{equation}  
Since the isospin breaking effects are almost negligible, we  
will consider only the $K^0\to\pi^-\nu{l}^+$ decay channel. Input  
values for the numerical calculations are given as follows:  
$m_K\simeq495$ MeV and $m_{\pi}\simeq140$ MeV,  
respectively.  The up- and down-quark masses are taken as their 
average value: $m_q\equiv(m_u+m_d)/2\simeq5$ MeV,   
while the strange-quark mass $m_s$ as around 150 MeV.    
  
It has been well-known that the experimental data for $f_{l+,0}$ can  
be reproduced qualitatively well by the linear and quadratic  
fits~\cite{Yao:2006px}:      
\begin{eqnarray}  
  \label{eq:ex}  
{\rm Linear}&:&\,f_{l+,0}(t)=f_{l+,0}(0)  
\left[1+\frac{\lambda_{l+,0}}{m^2_{\pi}}(t-m^2_l)  
\right],  
\nonumber\\  
{\rm Quadratic}&:&\,f_{l+,0}(t)  
=f_{l+,0}(0)\left[1+\frac{\lambda'_{l+,0}}{m^2_{\pi}}(t-m^2_l)  
+\frac{\lambda''_{l+,0}}{2m^4_{\pi}}(t-m^2_l)^2\right],  
\end{eqnarray}  
where $m_l$ is the lepton mass. The slope parameter $\lambda_{l+}$ is 
deeply related to the $K\to\pi$ decay radius ($\langle 
r^2\rangle^{K\pi}$) as follows~\cite{Bijnens:1994me}:    
\begin{equation}  
  \label{eq:cr}  
\lambda_+\simeq\frac{1}{6}\langle{r^2}\rangle^{K\pi}m^2_{\pi}.  
\end{equation}  
Moreover, this radius can be expressed in terms of the 
Gasser-Leutwyler low-energy constant $L_9$ in the large $N_c$   
limit~\cite{Gasser:1984ux}: 
\begin{equation}  
  \label{eq:gasser}  
L_9=\frac{1}{12}F^2_{\pi}\langle r^2 \rangle^{K\pi}.  
\end{equation}  
To obtain the decay rate $d\Gamma_{K\to l\nu\pi}$, we use the  
convention defined in Ref.~\cite{Bijnens:1994me}:  
\begin{equation}  
  \label{eq:dgamma1}  
  d\Gamma_{K\to l\nu\pi}=\frac{1}{16m_K(2\pi)^5}  
\sum_{\rm spins}\frac{d^3p_l}{E_l}  
\frac{d^3p_{\nu}}{E_{\nu}}\frac{d^3p_{\pi}}{E_{\pi}}  
\delta^4(p_K-p_l-p_{\nu}-p_{\pi})|T_{K\to l\nu\pi}|^2.  
\end{equation}  
This expression can be  further simplified as a function of  
$t$~\cite{Ji:2001pj}:   
\begin{eqnarray}  
  \label{eq:ratio}  
\frac{d\Gamma_{K\to l\nu\pi}}{dt}&=&\frac{G^2_F|V_{us}|^2}{24\pi^3}  
\left(1-\frac{m^2_l}{t}\right)^2   
\left[|\vec{p}_{\pi}|^3\left(1+\frac{m^2_l}{2t}\right)  
f^2_{l+}(t)+m^2_K|\vec{p}_{\pi}|\left(1-\frac{m^2_{\pi}}{m^2_K}\right)^2  
\frac{3m^2_l}{8t}f^2_{l0}(t)\right]  
\nonumber\\  
&\simeq&\frac{G^2_F|V_{us}|^2}{24\pi^3}|\vec{p}_{\pi}|^3f^2_{e+}(t)  
\,\,\,\,{\rm for}\,\,m_l=m_e\simeq0,   
\end{eqnarray}  
where the three momentum of the pion $|\vec{p}_{\pi}|$ is defined by:    
\begin{eqnarray}  
  \label{eq:pion}  
|\vec{p}_{\pi}|=\left[\left(\frac{m^2_K+m^2_{\pi}-t}{2m_K}\right)^2  
-m^2_{\pi}\right]^{\frac{1}{2}}.  
\end{eqnarray}  
This three momentum of the pion constrains the physically accessible  
region for the decay, i.e.:   
\begin{equation}  
  \label{eq:AR}  
  m^2_l\leq t\leq(m_K-m_{\pi})^2 . 
\end{equation}   
  
\section{Nonlocal chiral quark model from the instanton vacuum}  
In this section, we show how to derive the hadronic matrix element  
given in Eq.~(\ref{eq:DEF3}) within the framework of the nonlocal  
$\chi$QM from the instanton vacuum.  We begin by the low-energy   
effective QCD partition function derived from the instanton  
vacuum~\cite{Diakonov:1985eg,Musakhanov:1998wp,Musakhanov:2001pc,    
Musakhanov:2002vu}:  
\begin{eqnarray}    
{\cal Z}_{\rm eff.}&=&\int{\cal D}\psi{\cal D}\psi^\dagger   
{\cal D} {\cal M} \exp\int d^4 x\Big[ \psi^{\dagger}_{f} (x)    
(i\rlap{/}{\partial }+im_f)\psi_{f}(x)\nonumber    
\\&+&i\int\frac{d^4k\,d^4p}{(2\pi)^8} e^{-i(k-p)\cdot x}   
\psi^{\dagger}_{f}(k)\sqrt{M_f(k_{\mu})}   
U^{\gamma_5}_{fg} \sqrt{M_g(p_{\mu})}\psi_{g}(p)\Big].     
\label{effectiveaction}  
\end{eqnarray}   
$M_f(k)$ is  
the dynamically generated quark mass being momentum-dependent, whereas  
$m_f$ stands for the current-quark mass with flavor $f$. The nonlinear  
background pseudo-Goldstone field $U^{\gamma _{5}}$ is given by   
\begin{equation}  
U^{\gamma _{5}}=U(x)\frac{1+\gamma _{5}}{2}+U^{\dagger }(x)\frac{1-\gamma  
_{5}}{2}=1+\frac{i}{F_{\cal M}}\gamma _{5}\mathcal{M}\cdot  
\lambda-\frac{1}{2F_{\cal M}^{2}}(\mathcal{M}\cdot\lambda)^2\cdots  
\end{equation}   
with the meson decay constants $F_{\pi}=93$ MeV and $F_K=113$ MeV  
fixed to the experimental data.  The meson field $U$ is defined as  
$U=\exp[i\lambda\cdot\mathcal{M}/F_{\mathcal{M}}]$.  The octet  
pseudoscalar meson field $\mathcal{M}$ is defined as follows:     
\begin{eqnarray}  
  \label{eq:OT}  
\mathcal{M}\cdot \lambda&=&\sqrt{2}\left(  
    \begin{array}{ccc}  
\frac{1}{\sqrt{2}}\pi^0+\frac{1}{\sqrt{6}}\eta&\pi^+&K^+\\  
\pi^-&-\frac{1}{\sqrt{2}}\pi^0+\frac{1}{\sqrt{6}}\eta&K^0\\  
K^-&\bar{K}^0&-\frac{2}{\sqrt{6}}\eta\\  
    \end{array}  
\right),  
\end{eqnarray}  
  
Refs.~\cite{Musakhanov:1998wp,Musakhanov:2001pc} showed how to   
improve the low-energy effective QCD partition function in  
Eq.~(\ref{effectiveaction}) by taking into account effects of flavor  
SU(3) symmetry breaking, so that the dynamical quark mass acquires the  
contribution of the $m_f$ corrections:   
\begin{equation}   
M_{f}(k)=M_{0}F^2(k)\left[   
  \sqrt{1+\frac{m_{f}^{2}}{d^{2}}}-\frac{m_{f}}{d}\right],   
\label{eq:dqm}      
\end{equation}   
where $M_0$ is the dynamical quark mass with zero momentum transfer in   
the chiral limit.  Its value is determined by the saddle-point   
equation: $M_0\simeq350$ MeV. $F(k)$ is the momentum-dependent part   
which arises from the Fourier transform of the fermionic zero-mode   
solutions in the instantons.  However, we will employ  
the simple-pole type parameterization for $F(k)$:  
\begin{equation}  
  \label{eq:formfactor}  
F(k)=\frac{2\Lambda^2}{2\Lambda^2+k^2},\,\,\,\,  
\Lambda=\rho^{-1}\simeq600\,{\rm MeV}  
\end{equation}  
which shows a very similar behavior to the original expression of  
$F(k)$.  The value of $d$ in the square parenthesis of  
Eq.~(\ref{eq:dqm}) can be computed within the   
model~\cite{Pobylitsa:1989uq,Musakhanov:2001pc}:    
\begin{equation}  
  \label{eq:d}  
d=\sqrt{\frac{0.08385}{2N_c}}\frac{8\pi\bar{\rho}}{R^2}\simeq0.193\,{\rm GeV}.  
\end{equation}  
  
As mentioned previously, the momentum-dependent dynamical quark mass  
$M_f(k)$ breaks the conservation of the N\"other (vector) currents.  
Refs.~\cite{Musakhanov:2002xa,Kim:2004hd} derived   
the light-quark partition function in the presence of the external   
vector field:   
\begin{eqnarray}    
\tilde{{\cal Z}}_{\rm eff.}&=&\int{\cal D}\psi{\cal D}\psi^\dagger   
{\cal D} {\cal M}   
\exp\int d^4 x\Bigg[ \psi^{\dagger}_{f} (x)   
(i\rlap{/}{\partial}+\rlap{/}{V}+im_f)\psi_{f}(x)\nonumber   
\\&+&i\int\frac{d^4k\,d^4p}{(2\pi)^8} e^{-i(k-p)\cdot x}   
\psi^{\dagger}_{f}(k)\sqrt{M_f(k_{\mu}+V_{\mu})}   
U^{\gamma_5}_{fg}   
\sqrt{M_g(p_{\mu}+V_{\mu})}\psi_{g}(p)\Bigg].    
\label{GPF}   
\end{eqnarray}   
The effective chiral action then becomes:    
\begin{eqnarray}   
\label{eq:EA}   
\tilde{{\cal S}}_{\rm eff}=-N_c{\rm Tr}\ln\left[i\rlap{/}{\partial}   
+\rlap{/}{V}+im_f+i\sqrt{M_f(i\partial_{\mu}   
+V_{\mu})}   
U^{\gamma_5}_{fg}\sqrt{M_g(i\partial_{\mu}+V_{\mu})}\right],   
\end{eqnarray}   
where ${\rm Tr}$ denotes the trace over space-time, flavor, and spin  
spaces.  Calculating the functional derivative of   
$\tilde{{\cal S}}_{\rm eff}$ with the external vector field $V$, we  
obtain the relevant operator expression for the $K\to \pi$  
semileptonic decay form factors:   
\begin{eqnarray}  
  \label{eq:functional}  
&=&\left[\frac{\delta\tilde{{\cal S}}_{\rm eff.}}{\delta  
    V_{\mu}}\right]_{V=0} \nonumber\\  
&=& -N_c{\rm Tr}\left[  
\frac{1}{i\rlap{/}{\partial}+im_f+i\sqrt{M_f(i\partial_{\mu})}   
U^{\gamma_5}_{fg}\sqrt{M_g(i\partial_{\mu})}} \gamma_\mu\lambda^{4-i5}   
\right]_{V=0} \cr  
&&+N_c{\rm Tr}\left[  
\frac{\left[\frac{\partial}{\partial V_\mu}  
    \left(\sqrt{M_f(i\partial_{\mu})}\right)  
    U^{\gamma_5}_{fg}\sqrt{M_g(i\partial_{\mu})}   
-i\sqrt{M_g(i\partial_{\mu})}U^{\gamma_5}_{fg}  
\frac{\partial}{\partial V_\mu}\left(  
\sqrt{M_g(i\partial_{\mu})}\right)\right]\lambda^{4-i5}}   
{i\rlap{/}{\partial}+im_f+i\sqrt{M_f(i\partial_{\mu})}   
U^{\gamma_5}_{fg}\sqrt{M_g(i\partial_{\mu})}}  
\right]_{V=0}.  
\end{eqnarray}  
The first term in Eq.~(\ref{eq:functional}) is  
usually called a {\it local} contribution, and the other two terms  
the {\it nonlocal} ones.  Since we are interested in the decay process    
with two on-mass shell pseudoscalar mesons, that is, the pion and kaon  
as shown in Eq.~(\ref{eq:DEF3}), the local contribution can be written  
as follows:  
\begin{eqnarray}  
  \label{eq:expansion1}  
&&\left[\frac{\delta\tilde{{\cal S}}_{\rm eff.}}  
{\delta V_{\mu}}\right]^{K\pi}_{{\rm local},V=0}  
\nonumber\\  
&=&\frac{2N_c}{F_{\pi}F_K}{\rm Tr}  
\left[\frac{\sqrt{M_f(i\partial_{\mu})}   
\gamma_5{\cal M}^a\lambda^a\sqrt{M_g(i\partial_{\mu})}}  
{{\cal D}(i\partial_{\mu})}  
\frac{\gamma^{\mu}\lambda^{4-i5}}{{\cal D}(i\partial_{\mu})}  
\frac{\sqrt{M_f(i\partial_{\mu})}   
\gamma_5{\cal M}^b\lambda^b  
\sqrt{M_g(i\partial_{\mu})}}{{\cal D}(i\partial_{\mu})}  
\right].  
\end{eqnarray}   
The pseudoscalar meson field ${\cal{M}}^a$ can be either  
the kaon or the pion, depending on flavor.  $\mathcal{D}$ denotes   
the abbreviation for the quark-propagator:  
 \begin{equation}  
\label{eq:pro}  
{\cal D}_f(i\partial_{\mu})  
=i\rlap{/}{\partial}-i\left[m_f+M_f(i\partial_{\mu})\right].     
\end{equation}  
Then, the corresponding matrix element can be obtained as follows:  
\begin{equation}  
  \label{eq:mat}  
\Bigg\langle K(p_K)\Bigg|\left[\frac{\delta\tilde{{\cal S}}_{\rm eff.}}  
{\delta V_{\mu}}\right]^{K\pi}_{{\rm local},V=0}\Bigg|\pi(p_{\pi})  
\Bigg\rangle.  
\end{equation}  
The matrix element for the local contribution can be immediately  
expressed as  
\begin{eqnarray}  
  \label{eq:local}  
&&F^{\rm local(a)}_{\mu}=\frac{8N_c}{F_{\pi}F_K}\int\frac{d^4k}{(2\pi)^4}  
\frac{M_q(k_a)\sqrt{M_s(k_b)M_q(k_c)}}  
{\left[k^2_a+\overline{M}^2_q(k_a)\right]  
\left[k^2_b+\overline{M}^2_s(k_b)\right]  
\left[k^2_c+\overline{M}^2_q(k_c)\right]}  
\nonumber\\  
&&\hspace{2cm}\times  
\Bigg[\left[k_a\cdot{k_b}+\overline{M}_q(k_a)\overline{M}_s(k_b)\right]  
k_{c\mu}-\left[k_b\cdot{k_c}+\overline{M}_s(k_b)\overline{M}_q(k_c)\right]k_{a\mu}  
\nonumber\\  
&&\hspace{3cm}+  
\left[k_a\cdot{k_c}+\overline{M}_q(k_a)\overline{M}_q(k_c)\right]k_{b\mu}\Bigg],  
\end{eqnarray}  
where $\overline{M}_f(k)=m_f+M_f(k)$. The relevant momenta are defined as  
$k_a=k-p/2-q/2$, $k_b=k+p/2-q/2$ and $k_c=k+p/2+q/2$, in which $k$,  
$p$ and $q$ denote the internal quark, initial kaon, and  
transfered momenta, respectively, as depicted in   
Figure~\ref{fig0}.  The trace ${\rm tr}_{\gamma}$ runs   
over Dirac spin space.  Similarly, we can evaluate the nonlocal  
contributions as follows~\cite{Ryu:2006bf,Nam:2006sx}:     
\begin{eqnarray}   
\label{eq:nonlocal}  
&&F^{\rm nonlocal(b)}_{\mu}  
=\frac{8N_c}{F_{\pi}F_K}\int\frac{d^4k}{(2\pi)^4}  
\frac{\sqrt{M_q(k_c)}_{\mu}\sqrt{M_q(k_c)}M_q(k_a)M_s(k_b)}  
{\left[k^2_a+\overline{M}^2_q(k_a)\right]  
\left[k^2_b+\overline{M}^2_s(k_b)\right]  
\left[k^2_c+\overline{M}^2_q(k_c)\right]}  
\nonumber\\  
&&\times  
\left[\overline{M}_q(k_c)k_a\cdot k_b+\overline{M}_s(k_b)k_a\cdot  
  k_c-\overline{M}_q(k_a)k_b\cdot k_c  
+\overline{M}_q(k_a)\overline{M}_s(k_b)\overline{M}_q(k_c) \right]  
-\left(b\leftrightarrow c\right),  
\nonumber\\  
&&F^{\rm nonlocal(c)}_{\mu}  
=-\frac{4N_c}{F_{\pi}F_K}\int\frac{d^4k}{(2\pi)^4}  
\frac{  
\sqrt{M_q(k_a)}\sqrt{M_s(k_b)}\sqrt{M_q(k_c)}_{\mu}\sqrt{M_q(k_a)}  
\left[k_a\cdot k_b+\overline{M}_q(k_a)\overline{M}_s(k_b)\right]}  
{\left[k^2_a+\overline{M}^2_q(k_a)\right]\left[k^2_b+\overline{M}^2_s(k_b)\right]}  
\nonumber\\  
&&+\frac{4N_c}{F_{\pi}F_K}\int\frac{d^4k}{(2\pi)^4}  
\frac{\sqrt{M_q(k_a)}\sqrt{M_s(k_b)}\sqrt{M_q(k_c)}\sqrt{M_q(k_a)}_{\mu}  
\left[k_a\cdot k_b+\overline{M}_q(k_a)\overline{M}_s(k_b)\right]}  
{\left[k^2_a+\overline{M}^2_q(k_a)\right]\left[k^2_b+\overline{M}^2_s(k_b)\right]}  
\nonumber\\  
&&\hspace{4cm}+\left(b\leftrightarrow c\right),  
\end{eqnarray}  
where $\sqrt{M_f(k)}_{\mu}=\partial\sqrt{M_f(k)}/\partial k_{\mu}$.   
The local (a), nonlocal (b) and nonlocal (c) contributions correspond to the  
diagrams (a), (b) and (c) in Figure~\ref{fig0}, respectively.    
\section{Results and discussions}  
We now discuss various numerical results for the kaon semileptonic  
decay ($K_{l3}$) form factors in the present work.  We facilitate the   
Breit-momentum framework for the calculation, since we are free to 
choose an arbitrary momentum framework because of the Lorentz 
invariance.  The relevant momenta for the calculation are  
defined as follows ($-Q^2\equiv t>0$):    
\begin{eqnarray}  
\label{eq:k}  
p&=&\left(0,\,\,\,0,\,\,\,i\frac{\sqrt{t}}{2},\,\,\,  
\sqrt{\frac{m^2_K-m^2_{\pi}+t}{4\sqrt{t}}}  
\right),  
\,\,\,\,q=\left(0,\,\,\,0,\,\,\,-i\sqrt{t},\,\,\,0\right),  
\nonumber\\  
k&=&\left(k_r\sin\phi\sin\psi\cos\theta,\,\,\, 
k_r\sin\phi\sin\psi\sin\theta, 
\,\,\,k_r\sin\phi\cos\psi,\,\,\,k_r\cos\phi,\right).   
\end{eqnarray}  
  
\begin{figure}[t]  
\begin{center}  
\begin{tabular}{cc}  
\includegraphics[width=8cm]{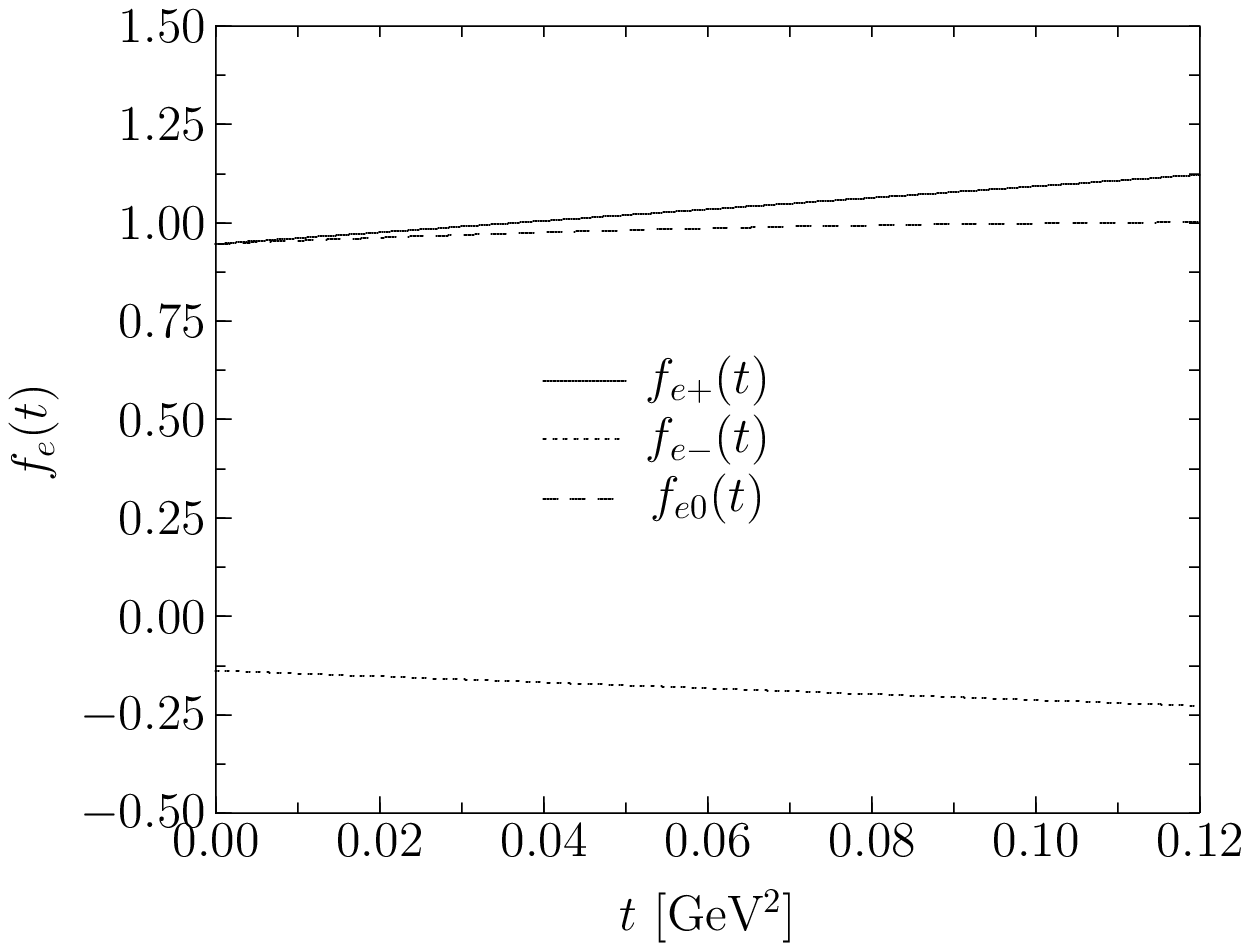}  
\includegraphics[width=8cm]{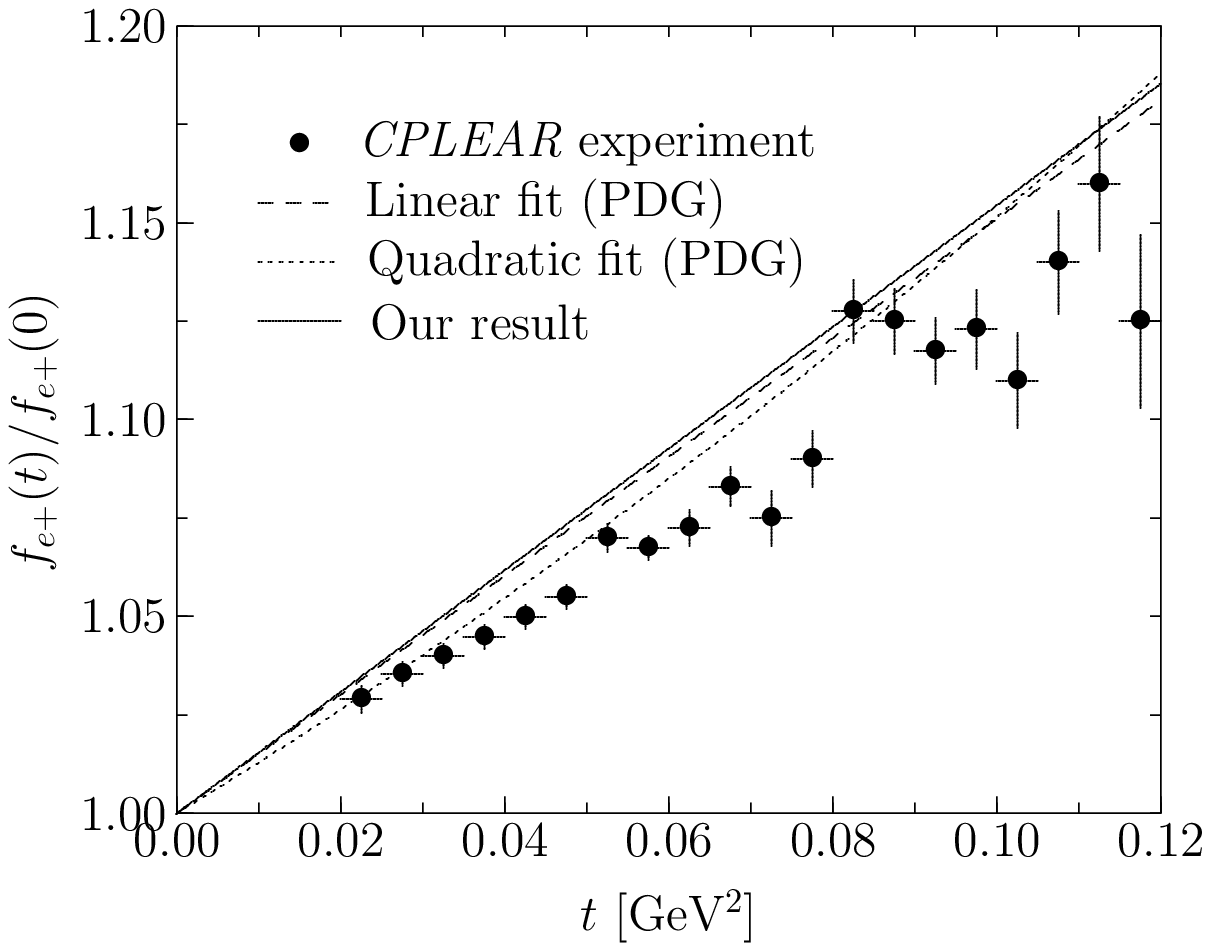}  
\end{tabular}  
\end{center}  
\caption{$K_{e3}$ form factors, $f_{e+}(t)$ (solid), $f_{e-}(t)$ (dotted) and  
$f_{e0}(t)$ (dashed) are shown in   the left panel, while in the right   
panel the ratio of $f_{e+}(t)$ and $f_{e+}(0)$ is given (solid).  We also   
draw the CPLEAR experimental data~\cite{Apostolakis:1999gs}, and linear  
(dashed) and quadratic (dotted) fits using the PDG  
data~\cite{Yao:2006px}.}           
\label{fig1}  
\end{figure}

We first consider the case of $K_{e3}$.  Since the electron mass is  
negligible in comparison to those of the pion and the kaon, it can be  
set to be zero.  In the left panel of Figure~\ref{fig1}, we draw the  
numerical results for $f_{e+}(t)$ (solid), $f_{e-}(t)$ (dotted) and  
$f_{e0}(t)$ (dashed) within the physically accessible regions  
constrained by Eq.~(\ref{eq:AR}).  Note that the scalar form factor  
$f_{e0}(t)$ is derived by using Eq.~(\ref{eq:swave}).    
We observe that the $f_{e+}(t)$ and $f_{e0}(t)$ are almost linearly  
increasing functions of $t$, whereas $f_{e-}(t)$ decreases. At $t=0$, our    
results demonstrate that $f_{e+}(0)=f_{e0}(0)=0.947$ and  
$f_{e-}(0)=-0.137$.  In the chiral limit, $f_{e+}(0)$ and $f_{e-}(0)$  
should be unity and zero, respectively, which is related to the      
Ademollo-Gatto theorem in the case of pseudo-Goldstone  
bosons~\cite{Ademollo:1964sr,Langacker:1973nf,Gasser:1984ux}:    
\begin{equation}  
\label{eq:GA}  
\lim_{q\to0}F^{\rm local(a)}_{\mu}\simeq 2p_{\mu} + \mathcal{O}(m_q).  
\end{equation}  
The Ademollo-Gatto theorem in Eq.~(\ref{eq:GA}) can be easily tested in the 
nonlocal $\chi$QM. Considering $q\to0$ and ignoring the terms being   
proportional to $k\cdot p$, we can rewrite the leading contribution of  
Eq.~(\ref{eq:local}) to order ${\cal O}(m_q)$ as follows:   
\begin{equation} 
  \label{eq:AG} 
\lim_{q\to0}F^{\rm local(a)}_{\mu}\simeq2\left[1+R(m_s)\right]p_{\mu}, 
\end{equation} 
where 
\begin{equation} 
  \label{eq:AG1} 
R(m_s)=\frac{1}{2}\left[\int\frac{d^4k}{(2\pi)^4}  
\frac{M^2(k)m_s\left[m_s+2M(k)\right]} 
{\left[k^2+M^2(k)\right]^3}\right] 
\left[\int\frac{d^4k}{(2\pi)^4}  
\frac{M^2(k)} 
{\left[k^2+M^2(k)\right]^2}\right]^{-1}.   
\end{equation} 
To evaluate Eq.~(\ref{eq:AG}), we employ the ratio $F_K/F_{\pi}$ 
computed within the same framework and expanded in terms of 
the strange quark mass ($m_s$) (see  
Refs.~\cite{Munczek:1988er,Munczek:1983dx,McKay:1983fm,McKay:1985mr,  
Atkinson:1990fn,Bowler:1994ir,Diakonov:1985eg} for more details):    
\begin{equation} 
  \label{eq:PS} 
\frac{F_K}{F_{\pi}}\simeq1+R(m_s). 
\end{equation}  
We also use that $k_b=k_c\to k+p/4$ since these 
two momenta share $p/2$ as   
$q\to0$. Note that we consider only the local 
contribution for $F_{\cal M}$ in   
Eq.~(\ref{eq:PS}). We, however, verified that   
the nonlocal contributions in Eq.~(\ref{eq:nonlocal}) also satisfy the 
Ademollo-Gatto theorem analytically.   
  
The effect of flavor SU(3) symmetry breaking is found to be  
rather small in the $K_{e3}$ form factor, i.e. its effect is around  
$5\,\%$.  In other approaches, for example, in $\chi$PT, the $K_{e3}$  
form factor is known to be $f_{e+}(t)=0.961\pm0.008$  
\cite{Leutwyler:1984je}, in LQCD,  
$f_{e+}(0)=0.960\pm0.009$~\cite{Becirevic:2004ya} and   
$0.952\pm0.006$~\cite{Tsutsui:2005cj}.     
  
In the right panel of Figure~\ref{fig1} we draw the ratio of  
$f_{e+}(t)$ and $f_{e+}(0)$ with respect to the CPLEAR experimental  
data~\cite{Apostolakis:1999gs}, and linear (dashed)  and quadratic (dotted)  
fits for the ratio using the PDG data~\cite{Yao:2006px}:  
$\lambda_{e+}=(2.960\pm0.05)\times10^{-2}$,  
$\lambda'_{e+}=(2.485\pm0.163)\times10^{-2}$, and     
$\lambda''_{e+}=(1.920\pm0.062)\times10^{-3}$.  In the present calculation,   
we obtain $\lambda_{e+}=3.028\times10^{-2}$ for the linear fit, which  
is very close to the experimental one, $2.960\times10^{-2}$.  Since  
our result for $f_{e+}$ is almost linear as shown in Fig.~\ref{fig0},  
we get almost a negligible value for the slope parameter $\lambda''$  
when the quadratic fit is taken into account. Being compared with   
other model calculations, the present results are comparable  
to those from $\chi$PT~\cite{Shabalin:1989cc,Bijnens:1994me}, and  
other models~\cite{Ji:2001pj,Kalinovsky:1996ii,Afanasev:1996my,  
Tsutsui:2005cj,Ji:2000fy}.  We compare explicitly the present results  
with those from other approaches in Table~\ref{table0}.    
  
Using Eq.~(\ref{eq:cr}) and Eq.~(\ref{eq:gasser}), we can easily  
estimate the $K_{e3}$ decay radius and low-energy constant $L_9$,    
respectively.  As for the $K_{e3}$ decay radius, we obtain   
$\langle{r}^2\rangle^{K\pi}=0.366\,{\rm{fm}}^2$.  This value is  
slightly larger than that in $\chi$PT~\cite{Gasser:1984ux} (see  
Table~\ref{table0}).  The low-energy constant $L_9$ turns out   
to be $6.78\times10^{-3}$, which is comparable to  
$7.1\sim7.4\times10^{-3}$~\cite{Gasser:1984ux} and   
$6.9\times10^{-3}$~\cite{Bijnens:1994me,Ecker:1994gg}.      
  
The ratio of the pion and kaon weak decay constants   
$F_K/F_{\pi}$ can be deduced from the scalar form factor $f_0$ via the  
Callan-Treiman soft-pion theorem~\cite{Callan:1966hu}. In the  
soft-pion limit ($p_{\pi}\to0$), the $K_{e3}$ form factor can be  
written as~\cite{lee:1968}:   
\begin{equation}  
\label{eq:CT2}  
\lim_{p_{\pi}\to0}F_{\mu}(p_{\pi},p_K)=p_{K\mu}\frac{F_K}{F_\pi}.    
\end{equation}  
Using Eqs.~(\ref{eq:DEF3}) and (\ref{eq:swave}), we obtain the  
following expression:  
\begin{equation}  
  \label{eq:CT1}  
\lim_{p_{\pi}\to0}F_{\mu}(p_{\pi},p_K)=  
\lim_{p_{\pi}\to0}(p_{\pi}+p_K)_{\mu}  
\left[f_{l+}(\Delta_{\rm CT})+f_{l-}(\Delta_{\rm CT})\right]  
\simeq p_{K\mu}f_{l0}(\Delta_{\rm CT}),  
\end{equation}  
where the value of $\Delta_{\rm CT}=m^2_K-m^2_{\pi}$ is called the  
Callan-Treiman point which can not be accessible physically.  Combining  
Eq.~(\ref{eq:CT2}) with Eq.(\ref{eq:CT1}), we finally arrive at  
the final expression of the $K_{l3}$ form factor for the  
Callan-Treiman theorem in terms of the scalar form factor and the  
ratio, $F_K/F_{\pi}$:     
\begin{equation}  
  \label{eq:CT}  
f_{e0}(\Delta_{\rm CT})=\frac{F_K}{F_{\pi}}.  
\end{equation}  
From our numerical calculation using Eq.~(\ref{eq:CT}), we find that   
$F_K/F_{\pi}=1.08$, which is around $10\,\%$ smaller than the 
empirical value ($1.22$).  The reason is due to the fact that the kaon 
weak decay constant turns out to be underestimated if we ignore the  
meson-loop $1/N_c$ corrections~\cite{Kim:2005jc}.  In the  
large $N_c$ limit the ratio can be expressed in terms of the 
low-energy constant $L_5$:  
\begin{equation}  
  \label{eq:chiral}  
\frac{F_K}{F_{\pi}}=1+\frac{4}{F^2_{\pi}}\left(m^2_K-m^2_{\pi}\right)L_5.      
\end{equation}  
Using the value of $F_K/F_{\pi}=1.08$, we obtain  
$L_5=7.67\times10^{-4}$ which is quite underestimated by about a 
factor 2, compared with the phenomenological value 
$1.4\times10^{-3}$~\cite{Ecker:1994gg}.  It is already well known that 
in order to reproduce the $L_5$ within the $\chi$QM the meson-loop 
$1/N_c$ corrections are essential.     
     
In the soft-pion limit, the model should satisfy the Callan-Treiman  
theorem given in Eq.~(\ref{eq:CT}).  Taking the limit $p_{\pi}\to0$  
for Eq.~(\ref{eq:local}), we can show that Eq.~(\ref{eq:local})   
satisfies the Callan-Treiman theorem, using Eq.~(\ref{eq:PS}) as follows:  
\begin{equation}  
  \label{eq:let}  
\lim_{p_{\pi}\to0} F^{\rm local(a)}_{\mu}\simeq\left[1+R(m_s)\right]p_{\mu},   
\end{equation}  
where $k_a= k_c\to k$ as $p_{\pi}\to 0$.  Inserting Eq.~(\ref{eq:PS})  
into Eq.~(\ref{eq:let}), we can show that the present result satisfies  
the Callan-Treiman theorem in Eq.~(\ref{eq:CT2}) (Eq.~(\ref{eq:CT})) 
in the case of the local contribution.  The nonlocal ones also fulfill 
the theorem. 
  
The decay width of  $K\to\pi\nu{e}$ can be easily computed by using  
the result of $f_{l+,0}$ and Eq.~(\ref{eq:ratio}).  It turns out that   
$\Gamma_{e3}=6.840\times10^6/$s and $\Gamma_{\mu3}=4.469\times10^6/$s  
with $|V_{us}|=0.22$ taken into  
account~\cite{Yao:2006px,Calderon:2001ni}. The results are slightly  
smaller than the experimental data ($\Gamma_{e3}=(7.920\pm0.040)\times10^6/$s  
and $\Gamma_{\mu3}=(5.285\pm0.024)\times10^6/$s)~\cite{Yao:2006px}.    
  
All numerical results are summarized in Table~\ref{table0} with  
the experimental data and those of other approaches for  
comparison.      
\begin{table}[ht]  
\begin{tabular}{c|c|c|c|c|c|c|c}  
\hline  
\hline  
$l=e(\mu)$&$f_{l+}(m^2_l)$&$-f_{l-}(m^2_l)$  
&$\lambda_{l+}\times10^2$&$\lambda_{l0}$&$\xi_l=|f_{l+}/f_{l-}|$  
&$\Gamma_{l3}$[$10^6/$s]  
&$\langle{r}^2\rangle^{K\pi}$[fm$^2$]\\    
\hline  
Present&0.947(0.963)&0.137(0.145)&3.03 &0.0136    
&0.147(0.152)&6.840(4.469)&0.366\\  
\cite{Chounet:1971yy}&$\cdots$&$\cdots$&2.45&$\cdots$&0.28&$\cdots$  
&0.292\\  
\cite{Gasser:1984ux}&1.022&$\cdots$&$\cdots$&$\cdots$&$\cdots$&$\cdots$  
&0.36\\  
\cite{Leutwyler:1984je}&0.972&$\cdots$&$\cdots$&$\cdots$&$\cdots$  
&$\cdots$&$\cdots$\\  
\cite{Tsutsui:2005cj}&0.952&$\cdots$&2.12&$\cdots$&$\cdots$&$\cdots$  
&0.376\\  
\cite{Ji:2001pj}&0.964&0.100&2.70&0.018&0.11&7.38(4.90)&0.322\\  
\cite{Kalinovsky:1996ii}&0.980(1.11)&0.24(0.27)&2.80&0.0026&0.35&$\cdots$  
&0.334\\  
\cite{Choi:1998jd}&0.962&$\cdots$&2.60&0.025&0.01&7.3(4.92)  
&0.310\\  
\cite{Afanasev:1996my}&$\cdots$&$\cdots$&2.80&$\cdots$&0.28&$\cdots$  
&0.334\\  
\cite{Scora:1995ty}&0.93&0.26&1.90&$\cdots$&0.28&$\cdots$  
&$\cdots$\\  
\cite{Cirigliano:2001mk}&0.9874&$\cdots$&$\cdots$&$\cdots$&$\cdots$  
&$\cdots$&$\cdots$\\  
\cite{Cirigliano:2004pv}&0.981&$\cdots$&$\cdots$&$\cdots$&$\cdots$  
&$\cdots$&$\cdots$\\   
\cite{Becirevic:2004bb}&0.960&$\cdots$&2.60&0.0089&$\cdots$&$\cdots$  
&0.310\\  
\cite{Barut:1969tm}&0.7&0.068&1.52&$\cdots$&0.097&$\cdots$&0.181\\  
\cite{Yao:2006px}(Exp.)&$\cdots$&$\cdots$&$2.96\pm0.05$&$\cdots$  
&$\cdots$&$7.920\pm0.040$&$\cdots$\\  
&&&&&&$(5.285\pm0.024)$&\\  
\cite{Apostolakis:1999gs}(Exp.)&$\cdots$&$\cdots$&$2.45\pm0.12$  
&$\cdots$&$\cdots$  
&$\cdots$&$\cdots$\\  
\hline  
\hline  
\end{tabular}  
\caption{Various numerical results for  
  $K^0\to\pi^-\nu{e}^+$($K^0\to\pi^-\nu{\mu}^+$). The   
  results from other model calculations and the experiments are listed as  
  well.}    
\label{table0}  
\end{table}     
  
\section{Summary and Conclusion}  
In the present work, we have investigated the kaon semileptonic decay  
($K_{l3}$) form factors within the framework of the gauged nonlocal  
chiral quark model from the instanton vacuum.  The effect of flavor  
SU(3) symmetry breaking were taken into account.  We calculated the  
vector form factors ($f_{\pm}$), scalar form factor ($f_0$),   
slope parameters ($\lambda_{+,0}$), decay width ($\Gamma_{l3}$), etc as  
demonstrated in Table~\ref{table0}.  We found that the present results  
of the kaon semileptonic decay form factors are in a qualitatively good  
agreement with experiments. We emphasize that there were no adjustable free  
parameters in the present investigation.  All results were obtained  
with only two parameters from the instanton vacuum, i.e. the average  
instanton size ($\bar{\rho}\sim1/3$ fm) and inter-instanton distance  
($R\sim1$ fm).   
  
In the present investigation, we have considered only the  
leading-order contributions in the large $N_c$ limit.  While these  
contributions reproduce the observables relevant for kaon semileptonic  
decay in general, it seems necessary to take into account the  
meson-loop $1/N_c$ corrections in order to reproduce quantitatively 
the kaon decay constant $f_K$ and the low-energy constant $L_5$.  As 
noticed already in Refs.~\cite{Nam:2006au,Gasser:1984gg,Kim:2005jc}, 
these meson-loop corrections can play an important role in producing   
the kaon properties.  The related works are under progress.  
\section*{Acknowledgments}   
The present work is supported by the Korea Research Foundation Grant  
funded by the Korean Government(MOEHRD) (KRF-2006-312-C00507). The work of  
S.i.N. is supported by the Brain Korea 21 (BK21) project in Center of 
Excellency for Developing Physics Researchers of Pusan National 
University, Korea. S.i.N. would like to appreciate the fruitful 
comments from M.~Khlopov.    
  
\end{document}